\documentclass[
reprint,
nobibnotes,
amsmath,amssymb,
 aps,
]{revtex4-1}

\usepackage{graphicx}% Include figure files
\usepackage{dcolumn}% Align table columns on decimal point
\usepackage{bm}% bold math
\begin{document}

\preprint{APS/123-QED}

\title{The effect of p-type doping on the oxidation of H-Si(111) studied by second-harmonic generation\\}% Force line breaks with \\

%%%%%%%%%%%%
 \author{Bilal Gokce}
 \author{Daniel B. Dougherty}
 \author{Kenan Gundogdu}
 \affiliation{%
 Physics Department, North Carolina State University, Raleigh NC 27695 \\}%

\date{\today}% It is always \today, today,

\begin{abstract}
Atomic force microscopy and second-harmonic generation data show that boron doping enhances the rate of oxidation of H-terminated silicon. Holes cause a greater increase in the reactivity of the Si-H up bonds than that of the Si-Si back bonds.
\begin{description}
\item[Usage]
Secondary publications and information retrieval purposes.
\item[PACS numbers]
 %\pacs{42.65.Ky} \packs{42.65.Ky} \packs{68.35.Dv} \packs{68.35.Gy}%\packs{68.35.Gy}
\end{description}
\end{abstract}

\pacs{Valid PACS appear here}% PACS, the Physics and Astronomy
\maketitle

%\tableofcontents
Surface chemical processes on Si, particularly oxidation, are important for many technological applications. In the Si-based microelectronics industry, SiO${_2}$ serves as gate dielectric. Therefore, the quality of the Si/SiO${_2}$ interface plays a significant role in the functionality of many Si based applications. Although most of the studies of the growth of SiO${_2}$ on Si have focused on thermal oxidation, native oxidation under ambient conditions is also important because during device production, H-passivated surfaces are often the starting point for sequences of chemical and lithographic processes. Therefore, the durability of H-passivated surfaces is a critical factor in device processing. 
 \\ \indent
In addition to the technologically important (001) surface, oxidation of (111) oriented surfaces has also been studied. These studies have provided a vast amount of information regarding surface chemistry of this fundamentally and technologically important chemical reaction. Availability of data from many different techniques makes (111) Si a convenient platform for studying surface chemical phenomena. For example IR absorption measures spectral changes in Si-H vibrations and relates them to chemical composition of the surface bonds.\cite{1,2} Ellipsometric measurements provide information about the thickness of the oxide.\cite{3,4} X-ray photoemission spectroscopy\cite{5,6} and photoelectron diffraction\cite{7} provide information about the suboxides at the Si/SiO${_2}$ interface. Reflectance difference spectroscopy measurements allow the number of oxidized Si layers to be counted during oxidation.\cite{8} Each of these techniques is sensitive to different aspects of chemistry.
 \\ \indent
A particularly powerful technique that has been used to study this interface is second harmonic generation (SHG). Since the dipole contribution is forbidden in centrosymmetric materials such as Si and amorphous materials such as SiO${_2}$, the Si/SiO${_2}$ interface provides the dominant contribution to the SHG signal in this system. Previously Daum et al. used spectrally resolved SHG to study suboxides and their evolution during Si oxidation.\cite{9,10} Downer and co-workers used SHG to probe excited hot carrier injection from Si-based films.\cite{11} Tolk and co-workers showed that boron-induced interface charge traps can be studied using SHG.\cite{12,13} In these studies the SHG response is typically obtained as the sample is rotated, or the sample is aligned so that the plane of incidence is always parallel to a certain crystal orientation. Similar measurements performed in ambient conditions on (111) Si have provided information about early oxidation kinetics and passivity of H-terminated surfaces.\cite{14,15}
\\ \indent
We recently employed SHG anisotropy (SHGA) measurements to study bond-specific oxidation of Si.\cite{16,17} By analyzing these data with the bond-charge model of nonlinear optics, we retrieve chemical kinetics characteristic of the different bonds.\cite{18} We discovered that tensile and compressive strain speed up and slow down, respectively, the oxidation kinetics of different bond directions, allowing control of in-plane surface chemistry during oxidation.\cite{19} In addition, we found that the excess electrons in a heavily n-type sample change oxidation kinetics of the different surface bonds dramatically.\cite{20} Here, we extend these studies to (111) surfaces of p-type Si wafers. We also performed atomic force microscopy (AFM) measurements, which provide additional information via changes in surface morphology. We find that p-type doping makes these surfaces very reactive to oxidation. The up-bonds are particularly affected, oxidizing very quickly and catalyzing the formation of the growing oxide layer. Despite the differences in chemical kinetics, the oxide terminates at the same thickness seen on n-type material.
\\ \indent
We investigated three (111) Si surfaces on boron-doped wafers with hole concentrations of 1.3 x 10$^{15}$, 3.6 x 10$^{16}$, and 5.1 x 10$^{18}$ cm$^{-3}$. These values were verified by four-point-probe experiments. The samples were Hyrdogen-passivated by the procedure described in Ref. 16 and then installed in an ambient atomic force microscope  within 3 minutes of passivation and imaged in non-contact mode as a function of time. 
\\ \indent
AFM experiments indicate significant surface reactivity in air. Figure 1 shows selections from the time dependent sequence of AFM images for lightly (upper panels) and heavily (lower panels) p-doped samples. In these sequences, the bright protrusions indicate the effect of surface oxidation as described previously.\cite{21} The effect of a high density of holes is evident by comparison. In panel a1 and b1, the initial H-passivated surfaces exhibit flat terraces separated by monatomic steps. For the lightly doped sample, oxidation takes longer than three hours. On the other hand, the heavily doped material oxidizes relatively rapidly. Most of the surface is covered with a structure consistent with a typical natural oxide in less than one hour. A high concentration of holes obviously impacts surface reactivity significantly.
\\ \indent
To gain further insight, we applied SHG to study the effect of p-type doping on the oxidation of the different surface bonds. The excitation beam was generated by a Ti-sapphire oscillator. It consisted of p-polarized 100 fs pulses centered at 806 nm arriving at a repetition rate of 70 MHz. The p-polarized SHG response was detected by a photomultiplier at every 1$^{\circ}$ as the sample was rotated by 360$^{\circ}$.
\\ \indent
Figures 2a and 2b show SHGA data for samples with hole concentrations of 3.6 x 10$^{16}$ and 5.1 x 10$^{18}$ cm$^{-3}$ respectively. Both are normalized to the level where the SHGA responses no longer change with time. According to the bond model\cite{18}, the three major features at azimuths 120$^{\circ}$, 0$^{\circ}$, and -120$^{\circ}$ arise when the alignment of the exciting field is roughly parallel to one of three back bonds. As reported previously\cite{16}, oxidation increases the magnitude of the SHGA signal at these azimuths by increasing bond asymmetry. A comparison of the two figures illustrates a significant dependence on carrier concentration. The three features of the heavily doped sample (Fig. 2b) rise almost linearly to their maximum value then remain constant thereafter. In contrast, the same three features of the moderately doped sample (Fig. 2a) reaches a higher maximum value and then decreases. We also note the existence of a 20 min incubation period for oxidation of the moderately doped sample.
\\ \indent
To obtain a more quantitative understanding of the reaction dynamics of oxidation, we investigated the time evolution of the hyperpolarizabilities for up and back bonds in more detail. The hyperpolarizabilities are determined by least-squares fitting Eq. (26) of ref. 18 to the SHGA data, as described in more detail there. Figure 3 summarizes the results for all three samples, thereby highlighting the differences that occur for different carrier concentrations. With increasing concentration the curves shift to earlier times. At the highest hole concentration investigated here, both the incubation period and the peak near 2h have vanished completely. This shows that the up bond is now oxidizing as rapidly as the back bonds. 
\\ \indent
We can make these observations quantitative by least-squares fitting simple exponential functions to the data. An expansion of the first 40 min of the data of Fig. 2a is shown in Fig. 4. The green curves show the exponential fits. The incubation period, if present, is fitted separately. The time constants for the incubation and rise periods are summarized in Table I. 

\begin{table}
\caption{Parameters yielding the best fits of single exponential functions to the data of Fig. 4.}
% title of Table
\centering
% used for centering table
\begin{tabular}{|c |c |c |c |c|}
% centered columns (5 columns)
\hline
& \multicolumn{4}{c|}{time constants [min]} \\[0.5ex]
\hline
& \multicolumn{2}{c|}{up bonds} &\multicolumn{2}{c|}{back bonds} \\ 
\hline 
&&&&\\ [-2ex]
p [cm$^{-3}$]& incubation & rise & incubation & rise \\ 
% inserts table
%heading
\hline 
&&&&\\ [-2ex]
% inserts single horizontal line
% inserting body of the table
1.3 x 10$^{15}$ & -5.2$\pm$0.5 & 33$\pm$2 & -23$\pm$2 & 15$\pm$2\\ 
3.6 x 10$^{16}$ & -1.2$\pm$0.3 & 28$\pm$2 & -22$\pm$2 & 13$\pm$2 \\ 
5.1 x 10$^{18}$ & fast & -22$\pm$4 & nonexistent & -54$\pm$4 \\ 
% [1ex] adds vertical space
\hline
%inserts single line
\end{tabular}
\label{table:nonlin}
% is used to refer this table in the text
\end{table} 

This analysis shows that the initial oxidation of p-type Si is dominated by the reaction dynamics of the up bonds and that holes facilitate the oxidation. For carrier concentrations of p = $\sim$ 10$^{15}$ cm$^{-3}$ as well as 10$^{16}$ cm$^{-3}$ we see that the incubation periods for the back bonds are identical. However, the reactivity of the up bonds clearly changes, highlighting the dominant influence of holes during the first step of oxidation, where the H of the Si-H up bond is replaced by -OH. Here, the up bonds get through their incubation phase faster, the smaller negative incubation time constants of the up bonds compared to the back bonds indicate that up bonds reach saturation quicker than the back bonds. But their rise time constants are considerably slower than that of the back bonds once the back bonds start to oxidize as time evolves and -OH caps become available in the second step of oxidation. At this time the back bonds become more reactive and dominate the oxidation process with twice the rate of the up bonds. This is reflected in the rise times for these bonds. Positive time constants refer to new processes becoming active, the smaller these numbers get the faster the process is becoming active. For back bonds we obtain small positive time constants showing the dominance of these bonds once they become active. The extreme case is the p = 5.1 x 10$^{18}$ cm$^{-3}$ sample. Here the incubation appears to be completely missing and the evolution is a simple exponential rise with a negative time constant, suggesting that the oxidation process is rising to a saturation level. Our ellipsometric data (not shown) show that for both n- and p-type samples oxidation continues until the thickness reaches a certain value, that is independent of carrier type and concentration\cite{20}. These results are consistent with corrosion studies of Si and Ge in aqueous solutions, where the availability of holes which weakens the surface bonds is the rate-limiting step.\cite{22,23}
\\ \indent
In conclusion, our AFM and SHGA data show that presence of holes in p-type material significantly increases the oxidation rates of H-terminated (111) Si samples, with the up bonds oxidizing at least as fast as the back bonds, and probably faster.  \newline

\textit{The authors acknowledge D. E. Aspnes for fruitful discussions and J. Reynolds for the resistivity measurements of our samples.}

% Produces the bibliography via BibTeX.

\section{Figure Captions}
FIG. 1. AFM scans of boron doped (111) Si samples for 0h, 1h, and 3h of oxidation in air 
(a1)-(a3) carrier concentration p=1.3 x 10$^{15}$ cm$^{-3}$ (b1)-(b3) carrier concentration p=5.1 x 10$^{18}$ cm$^{-3}$ 

FIG. 2. (a) Evolution of SHGA during air exposure of an H-terminated (111) Si sample with p=3.6 x 10$^{16}$ cm$^{-3}$
(b) as (a) but for p=5.1 x 10$^{18}$ cm$^{-3}$

FIG. 3. (a) Evolution of the average hyperpolarizabilities of the up bonds for boron-doped (111) Si samples with p = 1.3 x 10$^{15}$, 3.6 x 10$^{16}$, and 5.1 x 10$^{18}$ cm$^{-3}$ .  (b) As (a), but for the Si back bonds

FIG. 4. (a) first 40 min of Fig 3a; (b) first 40 min of Fig 3b.   The green curves show the exponential fits.
 
\begin{figure}[h]
\includegraphics[scale = 0.4]{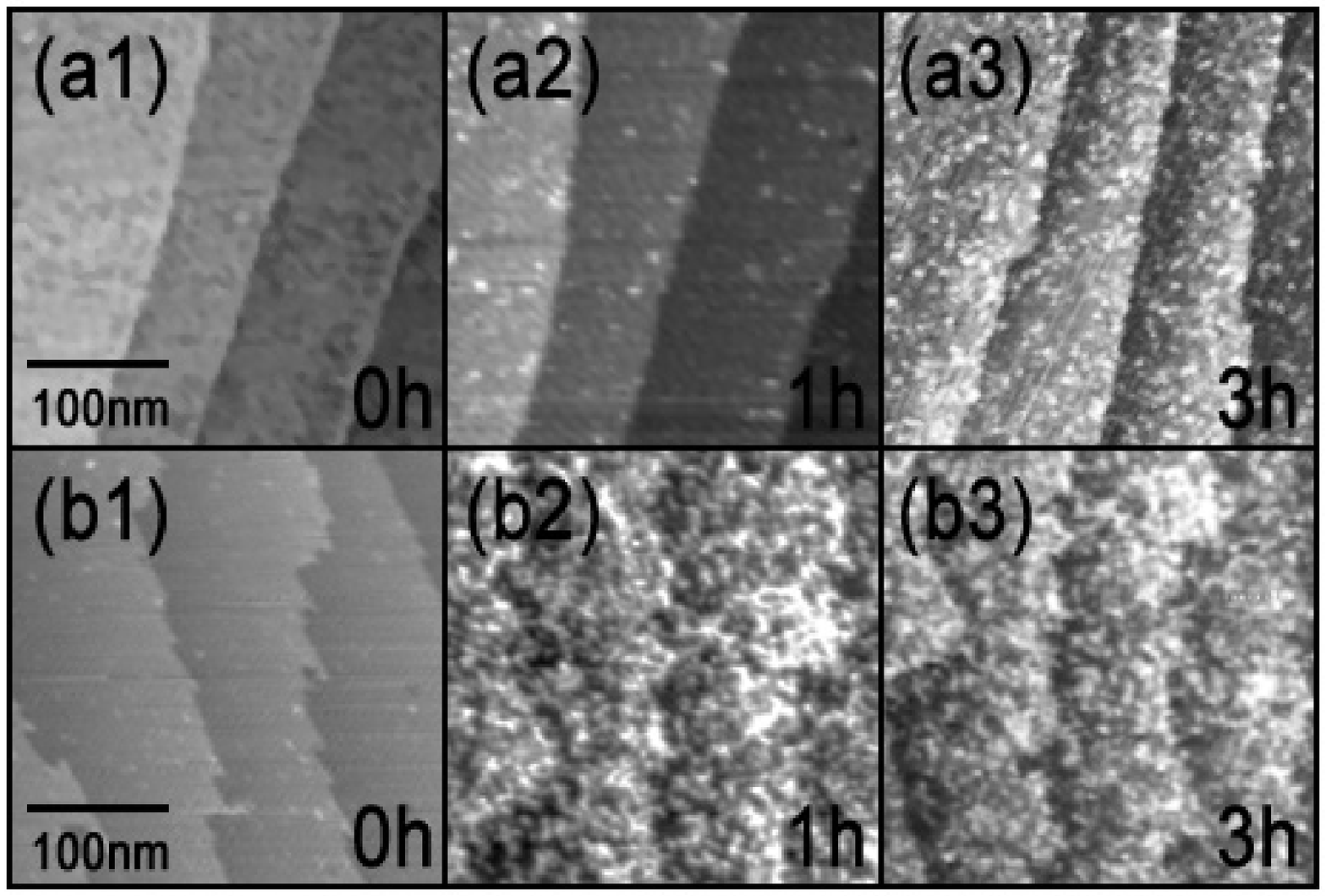}% Here is how to import EPS art
\caption{\label{fig:epsart} }
\end{figure}

\begin{figure}[h]
\includegraphics[scale = .127]{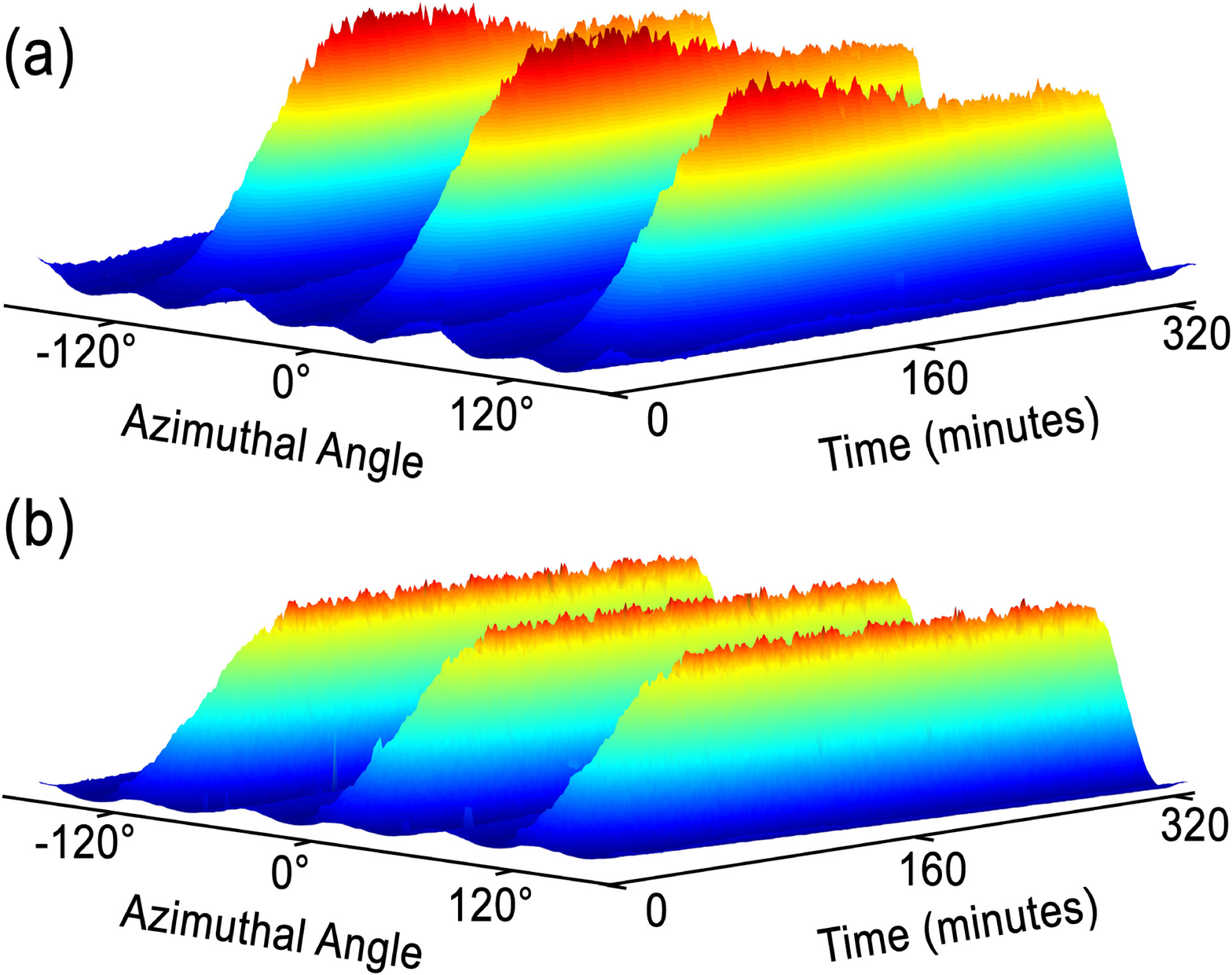}% Here is how to import EPS art .127 
\caption{\label{fig:epsart2} }
\end{figure}

\begin{figure}[h]
\includegraphics[scale = .127]{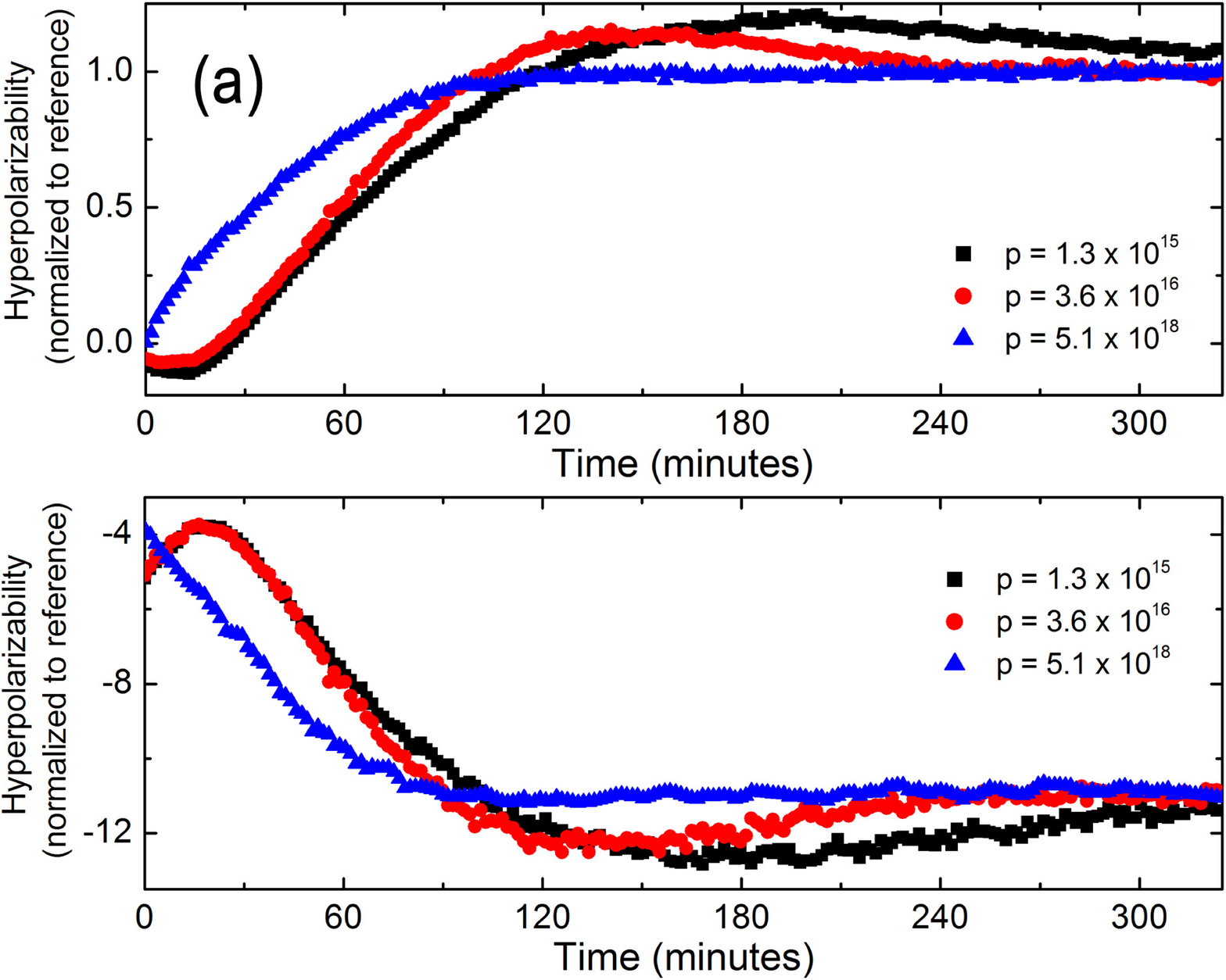}% Here is how to import EPS art
\caption{\label{fig:epsart3}  }
\end{figure}

\begin{figure}[h]
\includegraphics[scale = .127]{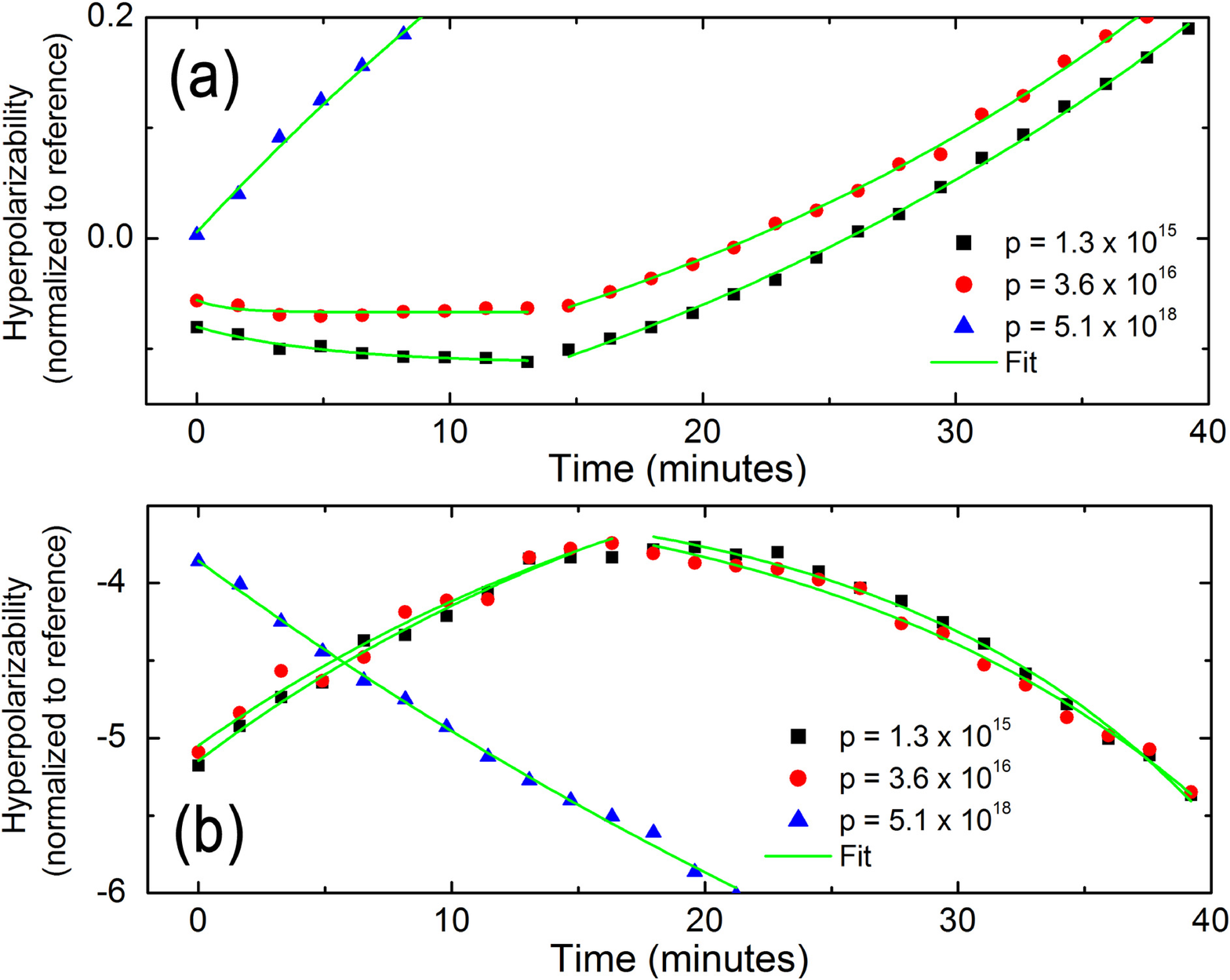}% Here is how to import EPS art
\caption{\label{fig:epsart4}  }
\end{figure}

\end{document}